# Haze at Occator crater on dwarf planet Ceres

G. Thangjam[1], M. Hoffmann[1], A. Nathues[1], J.-Y. Li[2], and T. Platz[1]

[1]Max Planck Institute for Solar System Research, Justus-von-Liebig-Weg 3, 37077 Goettingen, Germany, (thangjam@mps.mpg.de, hoffmann@mps.mpg.de, nathues@mps.mpg.de, platz@mps.mpg.de)

[2]Planetary Science Institute, 1700 East Fort Lowell Rd, Suite 106, Tucson, AZ 85719-2395, USA (jyli@psi.edu)

Corresponding author: thangjam@mps.mpg.de

**Abstract**

A diurnal varying haze layer at the bright spots of Occator on dwarf planet Ceres has been reported from images of the Dawn Framing Camera (Nathues et al. 2015). This finding is supported by ground-based observations revealing diurnal albedo changes at Occator's longitude (Molaro et al. 2016). In the present work, we further investigate the previously reported haze phenomenon in more detail using additional Framing Camera images. We demonstrate that the light scattering behavior at the central floor of Occator is different compared to a typical cerean surface and is likely inconsistent with a pure solid surface scatterer. The identified deviation is best explained by an additional component to the scattered light of the surface, i.e., a haze layer. Our results support the water vapor detection by Herschel observations (Küppers et al. 2014) though the existence of a tenuous cerean exosphere is not yet confirmed (Roth et al. 2016).

**Introduction**

Dwarf planet Ceres (diameter ~960 km) is the largest object in the main asteroid belt located at a mean heliocentric distance of ~2.8 AU (Russell et al. 2016). The physical characteristics (e.g., shape, density, mass) constrain a differentiated and thermally evolved volatile-rich body (McCord and Sotin 2005; Castillo-Rogez and McCord 2010; Park et al. 2016). Ceres is a 'primitive' object that might hold clues to understand the 'snow line' or 'water-frost line' in the solar system (Nathues et al. 2015; Rivkin et al. 2016). Recently, Nathues et al. (2015) reported a near-surface haze, filling a part of the Occator crater using images acquired at oblique views from the Framing Camera (FC) onboard NASA's Dawn spacecraft. They found an unexpected brightness excess at the central floor region changing on a diurnal basis. Nathues et al. (2015) concluded that the brightness excess is the result of a localized haze layer forming by (water-ice) sublimation at the bright spots. Similarly, diurnal albedo variability at the longitude that corresponds to the Occator bright spots was observed by means of ground-based radial velocity changes (Molaro et al. 2016). These observations support the water vapor detection by the Herschel telescope (Küppers et al. 2014). However, remote detection and existence of a tenuous $H_2O$ exosphere on Ceres is still under debate (A'Hearn and Feldman 1992; Küppers et al. 2014; Roth et al. 2016). The occurrence of haze (particles) requires a volatile driving phase that could be water ice, whose presence is up to now unconfirmed by the Visible and Infrared Spectrometer (VIR) onboard Dawn; though a small amount of water ice (~1 vol. %) is spectrally compatible (De Sanctis et al. 2016). However, the spectral detection of water-ice at Oxo crater (Combe et al. 2016) supports in general the haze explanation since Oxo is the second site for which haze has been reported by Nathues et al. (2015).

In the following we present a more detailed analysis of the haze phenomenon and the results that led to this conclusion. We also address complications and limitations in characterizing optical and scattering properties of the near-surface haze.

**Data and Methodology**

Clear filter (panchromatic) FC images with a large bandwidth between ~450 nm and ~920 nm (Sierks et al. 2011) are used in our analysis. The images were obtained from ~14,000 km distance during RC3 (Rotational Characterization 3) north (images FC21B00/36586 – 36681, obtained on 04-05-2015, and 'RC3N' hereafter) and south cycle (FC21B00/37104 – 37169, obtained on 07-05-2015, and 'RC3S' hereafter). The spatial resolution is ~1.2 km/pixel. These images were selected since they have been obtained under a broad range of emission (70°-22°-79° for RC3N, 78°-60°-77° for RC3S) and incidence (84°-16°-82° for RC3N, 63°-16°-47° for RC3S) angles, with a narrow range of phase angles (35-39° for RC3N and 42-45° for RC3S) (cf. extended data figure 2 in Platz et al. accepted). The RC3N images cover the entire local day while the RC3S covers less. First, our analysis concentrates on RC3N and is then extended to RC3S. Orbit data showing higher spatial resolution could not be used since these data are taken exclusively at nadir or close to nadir pointing, and therefore they do not provide the required observational geometry for our study.

We used FC level 1b images (radiometrically calibrated data) with 14 bit resolution (Nathues et al. 2014, 2015). Data processing and retrieval of required information, as for example, the observation geometry are performed using the Integrated Software for Imagers and Spectrometers/ISIS software (Anderson et al. 2004). Figure 1 shows Occator crater at local morning, noon and evening from a polar orbit view (RC3N) displaying the corresponding illuminated Ceres disk and the location of Occator (inset subfigures). In Fig. 1 right panels show the measured flux profiles [W/(m$^2$Sr)] versus spatial pixels along the trace (x-y) marked in the left panels. We used images in a body-centered coordinate system since any map projection of the near limb views of Occator would lead to significant pixel distortions. The central bright spot is termed 'primary', the cluster towards the east of the primary is termed 'secondary', and the space between the primary and secondary area is the 'bridge' (Fig. 1). Profiles across the bright spots are drawn passing through their centers defined by their respective maximum flux values (Fig. 1). Figure 2 shows the corresponding views of RC3S data, which are characterized by a diurnal geometry leading to much higher emission angles around noon. All images in Figs. 1 and 2 are at similar local times and are displayed with the same image stretch. Since the primary spot is saturated in many of the FC images, the center of the widest areal extension of the saturated region at the primary spot was used to define the trace through the primary spot. The positioning of the trace from one image to the next required particular attention since slight offsets in the order of a few pixels can yield to different flux values.

In order to analyze the unusual photometric phenomena at Occator (Nathues et al. 2015), we had to assume an unknown distribution and structure of the haze inside the crater, which does not allow a straight comparison of individual fluxes and thus we did not perform a photometric correction. However, the ratios of fluxes at specific locations inside Occator are suitable to distinguish influences of haze from pure surface scattering, in particular, if compared with a reference area. The secondary spots are not saturated unlike the primary spot, and they are less affected by rough topography compared to the primary spots and thus less influenced by shadows, especially for high solar incidence angles. Therefore, the secondary spot is a suitable site that can be used for our investigation ('secondary' hereafter). A further suitable target is the 'bridge', since this surface area is expected to be influenced by activities at the bright spots, but on a lower scale. Thus we used the flux measured at the secondary spot closest to the central spot (maximum) and the bridge (minimum) for our ratio (see Figs. 1, 2), which was plotted versus the observational geometric parameters. Detection of an unusual photometric response is possible by comparison of the geometric dependence of the flux ratio at Occator and a ratio at a reference area (i.e., without the haze component). In the following we compare the geometric dependence of the Occator flux ratio with results of a typical solid surface scatterer outside Occator. As references we chose sites at a medium-sized crater (~38 km diameter, 14.2° N, 256° E) located more than 100 km north-west of Occator which we assumed to show a typical solid surface scattering property (Fig. 3); here we selected a brighter and a dark site similar as at Occator. The fluxes at the reference sites are measured using RC3N images. The respective ratio is plotted versus illumination and viewing geometry parameters in Fig. 3 and compared with those of Occator's central floor (Figs. 4 and 5). Differences between the data at the target and reference area will be crucial to understand the scattering behavior at the spots of Occator and an additional haze layer, e.g. the relevance to optical depths and changes of its structure during the local time. Specifically, we are going to test the ratios of the fluxes of the secondary spot and the dark surface) from noon to dawn and dusk at different observation geometries, i.e.,

$$\frac{F_{bn}/F_{dn}}{F_{bd}/F_{dd}}$$

where, $F_{bn}$ and $F_{dn}$ are the flux at noon for the secondary and the bridge, and $F_{bd}$ and $F_{dd}$ are the flux at dawn or dusk for the secondary and the bridge, respectively.

In order to understand whether the higher multiple scattering proportion of the bright secondary could be responsible of the ratio increase shown in Fig. 4 we estimated its contribution to our data. In models describing atmospheric haze the contribution from multiple scattering is significant (Hansen 1969) as it is for bright solid surface material, i.e., reflectances >> 0.02 but < 0.95 or "translucent" materials (Buratti and Veverka 1985; Holzschuch and Gascuel 2013). Recent high spatial resolution (~35 m/pix) datasets show reflectance values > 0.3 (bidirectional reflectance with respect to a standard viewing geometry using Hapke functions, i.e., 30° incidence and 0° emission angle) for the bright material on Ceres (Nathues et al. submitted). Low resolution FC RC3 data, however, show reflectances of bright material (primary spot) and global surface of ~0.25 and ~0.03 at 0.55 µm, respectively (Nathues et al. 2015). However, the primary spot is saturated in many of our images, and thus we used a secondary spot for our study. This spot shows a reflectance of about 3-4 times of the global surface. Thus we can estimate the contribution of the multiple scattering components. Considering an isotropic single-scattering phase function (Hapke 2012; Li et al., 2006, 2013, 2016) and single scattering albedo from Li et al. (2016), we found that the multiple scattering can contribute upto ~ 12%, 30% and 2% for the secondary spot, bright spot and the average Ceres surface, respectively. We also computed the multiple scattering component for extreme illumination angles in our data assuming an isotropic particle scattering, i.e., $p(\alpha) \sim 1$ in the Hapke model of the bidirectional reflectance of a particulate medium $[p(\alpha)+M(i,e,w)]$, where $p(\alpha)$ is the average particle scattering function, and $M(i,e,w)$ is the multiple scattering component, and $w$ is the single scattering albedo.

For RC3N data, $M(i,e,w)$ for the dark surface is 0.03 (i~84, e~70) and 0.07 (i~16, e~22), while it is 0.15 (i~84, e~70) and 0.39 (i~16, e~22) for the secondary spot. Similarly, for RC3S, it is 0.04 (i~63, e~78) and 0.06 (i~16, e~60) for the dark surface, while it is 0.20 (i~63, e~78) and 0.33 (i~16, e~60) for the secondary spot. In these cases, $M(i,e,w)$ is far less than 1, and it implies that multiple scattering is negligible in this analysis, though it is expected to be present as a minor component.

In order to test our results and to decide, if multiple scattering has a significant influence on them, we calculate the diurnal effect on the ratios mentioned above. These contributions are 1.11 for RC3N and 1.10 for RC3S. Any observed change in the ratios exceeding this contribution significantly has to be explained beyond multiple scattering.

Statistical errors introduced by the signal-to-noise ratio of the images and resulting error propagation for the calculated ratios have been determined (Sierks et al. 2011; Thangjam et al. 2014). Uncertainties are usually too small to be represented in our figures. However, unidentified systematic errors may be present, which are responsible for the obvious scatter in our plots, possibly caused by sub-pixel influence of shadows and from local small-scale topography. Then, these two sources of errors are added in quadrature as is shown as error bars in Figs. 3, 4, and 5.

**Results**

The flux ratio of the selected two reference sites is constant (see Fig. 3) when plotted against several geometric parameters, i.e. cosine of the incidence angle (cos i), cosine of the emission angle (cos e), cosine of phase angle (cos α), and a geometric factor that describes a first order approximation to penetration of incident light and scattered light on the surface [cos i / (cos i + cos e)]. Constant trends as shown in Fig. 3 are expected for a solid surface scatterer since we can assume that scattering follows a standard photometric model (e.g., Lommel-Seeliger, Hapke). In this case the flux ratio at two locations (assuming similar scattering parameters) depends only on the ratio of their geometric albedos.

Figure 4 displays the flux ratios for RC3N data between the secondary and the bridge versus the geometric parameters. We found an increasing ratio starting at ~1.8 up to ~3 that corresponds to the decreasing trend of incidence and emission angles from dawn or dusk to noon (Fig. 4A-D). These trends are clearly different compared to the constant trends of the reference sites as shown in Fig. 3. In addition, Fig. 4 displays the flux ratio versus flux at the secondary and its diurnal variation in local solar time. The flux ratio trend from local morning to noon and noon to evening differs slightly, indicating a diurnal dichotomy (Fig. 4D/F). While the diurnal dichotomy is small for data describing the dependence on incidence and emission angle (Fig. 4A/B), the dichotomy is more pronounced in Figs. 5D/F. However, Figure 4C and 5C demonstrate the non-correlation between the flux ratio and the phase angle, even despite different trends of the phase angle.

In contrast to the ratio of our reference sites the ratio at Occator changes linearly with the signal at the secondary bright spot (Fig. 4E). Figure 4F displays the dependence of the ratio on the local time and shows a slightly asymmetric pattern with its maximum at noon (note the slight slope differences before and after noon that are stronger pronounced in Fig 5F). Figure 5 shows the corresponding trends of the Occator sites during RC3S but now for higher emission angles. While the diurnal asymmetry is here more pronounced than in Fig. 4D/F, the ratio range is lower, ranging from 1.6 to 2.4. This shows that the contrast (expressed by the ratio) is lowered for the higher emission angles in RC3S data. In other words, the lower contrast at noon appears to be influenced by a larger optical depth in the RC3S data compared to the RC3N data.

The results obtained from RC3N (Fig. 4) and RC3S (Fig. 5) at Occator are similar but differ clearly from the trends exhibited by the reference area (Fig. 3). Such a deviation from a solid scatterer can be explained by an additional component, i.e., a semi-transparent layer of haze as suggested by Nathues et al. (2015).

A layer of haze, however, above a solid surface can be expected to show variable extent and a non-homogeneous optical thickness at different illuminations. Combined with different emission angles this leads to a variable optical depth influencing the balance of scattered and transmitted light. In a haze layer, for instance, a small decrease of transmission may be combined with a considerable increase in the flux level, and a corresponding decrease of the contrast. Thus the measured flux shows quite unusual properties compared with the usual dependence on the phase angle alone.

There is a relative decrease of optical path length with the viewing direction in Figs. 1B and 2B (noon) and the direction of the profile through the bridge and secondary spot, which furthermore have opposite orientations in RC3N and RC3S views. This influences the ratio of morning/evening data relative to those of noon data. This and the different contribution from effects by incidence and emission angles seem to be responsible for the inversion of trends in the diurnal dependence (Figs. 4D and 5D). The non-linear trends of the ratios from dawn or dusk to noon in Figs. 4 and 5 also include information on the individual contribution from the diurnal change of the degree of activity leading to the haze by insolation and also due to viewing effects. If the dependence on the emission angle is nullified, a part of the trends can only be explained by such a diurnal variability of the source. Thus there is a simultaneous change of optical depth due to optical thickness, the geometric structure, and the viewing geometry of the haze, which is responsible to the deviation from a constant ratio, independent from the investigated parameters.

**Discussion**

The results and conclusion made by Nathues et al. (2015) using level 1a data (uncalibrated data in DN values ranging from 1 to 16383) are confirmed by this study using level 1b data (radiometrically corrected flux data) from RC3N and RC3S. Albedo variation on a diurnal basis at Occator longitude is also reported by ground-based observation using radial velocity measurements (Molaro et al. 2015). Notably their observation was made independently to the Nathues et al. (2015) finding. These studies support the water vapor detection by Küppers et al. (2014) using Herschel space-based telescopic data. The existence of haze is further supported by the spectral detection of water-ice in Oxo crater (Combe et al. 2016). As mentioned previously, Nathues et al. (submitted) reported that Occator and Oxo ('feature A' in their publication) do exhibit unusual scattering behavior and haze.

Note that the data used in our analysis are not photometrically corrected since the photometric parameters of surface and haze cannot be derived with the required precision. As a starting point to describe surface scattering, the Lambert model can be used, considering only the dependence of the scattering by the cosine of the incidence angle. By adding further parameters, for example, the emission angle (Lommel-Seeliger) or backward to forward scatter ratio and roughness (Hapke 1993) improved representations of an optically thick planetary surface can be modeled. However, in case of a solid surface superposed by a haze layer, these models are inadequate. Models which consider the cumulative semi-transparency of a finite scattering layer (e.g., Blinn 1982) seem appropriate, if conditions are spatially and temporary constant. Scattering in a haze layer by particles of unknown reflectance in a multi-component source (haze plus surface) implies scattering and absorption. Both may be present but cannot be identified and separated by our data. Furthermore, it cannot be excluded that a minor part of the phenomena may be related to non-uniform scattering depending on the azimuth, since deposition of the haze particles may have occurred preferentially on slopes facing the source. Since we have also to consider the possibility of a non-uniform deposition of the bright material, this may cause a correlation between reflectance and azimuth. Moreover, a diurnal change may imply variable sizes and density of haze particles and implicitly variable shape of the haze layer, which leaves an accurate photometric modeling too complex to achieve for single images.

The main conclusion from the data shown in figs. 3 to 5 is an increase of 50% of the ratio of reflectances of bright to dark in RC3N (Fig. 4) and also 50% in RC3S (Fig. 5), contrasting to what is measured at the reference area (Fig. 3). Thus the large increase of the flux ratio (in Figs. 4 and 5) from high to low incidence angles exceeds our expected influence by multiple scattering about four to five times and thus cannot be responsible for the unusual trend we observed. In fact, the effect is so large that its majority must be explained by a further component.

We also like to point out that optical vignetting or a potential contamination by instrumental stray light is negligible (Sierks et al. 2011) compared with the observed local contrasts and our results and conclusions, because the point spread function of the pixels (FWHM 1.7px) do not play a significant role for our analysis, despite the remarkable surface contrasts we have to deal with (Nathues et al., 2016).

Due to the panchromatic measurements we are not able to derive information on the composition of the haze, but, a high similarity to the materials detected as deposit on the nearby surface seems likely. The surface material at Occator consists mainly of dark carbonaceous material mixed with ammonia-bearing phyllosilicates and (possibly) carbonates (De Sanctis et al. 2016). There are compositional variations from the bright central dome of Occator to the less bright spots on its floor and the dark floor material (De Sanctis et al. 2016; Nathues et al. submitted). The haze may consist of condensed volatiles and solid particles lifted by sublimation. The possibility of haze at crater Oxo is strengthened by recent VIR spectroscopic observations which detected water-ice (Combe et al. 2016). Also the spectral information obtained at Occator does not rule out the presence of water ice (De Sanctis et al., 2016). Note that De Sanctis et al. (2016) reported a small amount of water ice in Occator (~ 1 vol. %) despite the detection limits of VIR instrument (coarse spatial resolution and the signal to noise constraints). The sublimating ice is possibly nested in fractures or the haze particles are entrained by gases, which are likely impossible to detect by Dawn's payload (Nathues et al. submitted).

**Conclusion**

The diurnal brightness excess variation at Occator's bright spots as found by Nathues et al. (2015) is confirmed. At Occator's floor the scattering properties differ from the standard cerean surface. Scattering models and lab measurements need to show whether the haze explanation remains the only valid one. At the moment this scattering behavior can be best explained by a combined effect of a solid surface scatterer and a haze layer. The available data are insufficient for an analysis of the optical properties of this haze, but it is most likely an optically thin and semi-transparent layer forming at near-surface level. Diurnal albedo variations that correspond to Occator's longitude have been detected by radial velocity changes (Molaro et al. 2016) supporting the existence of a temporarily varying haze layer. The detection of water-ice signatures at Oxo further supports ongoing sublimation activities on Ceres. Although we expect the maximum haze concentration above the central spot in Occator, it is remarkable that our data also indicate activity at the secondary spots. Besides effects caused by geometric parameters, as shown by our data, the diurnal variability of the haze implies a changing composition or microphysical or optical properties of the haze, e.g., a change of the distribution of particle sizes, density, shape, and thus making it difficult to optically characterize the haze layer. Data from the extended mission at Ceres closer to perihelion may provide an opportunity to detect and characterize the haze phenomenon and its origin.

**Acknowledgement**

We thank the Dawn mission's operations team for the development, cruise, orbital insertion, and operations of the Dawn spacecraft at Ceres, and also the Framing Camera operations team, especially P. G. Gutierrez-Marques, I. Hall and I. Büttner. The Framing Camera project is financially supported by the Max Planck Society and the German Space Agency, DLR.

**Reference**

A'Hearn, M. F., Feldman, P. D. 1992, *Icar* 98, 54

Anderson, J.A., Sides, S.C., Soltesz, D. L., et al. 2004, LPSC 35, abstract 2039

Blinn, J. F., 1982. *Comput. Graph.* 16, 21

Buratti, B. J., Veverka, J. 1985, *Icar* 64, 320

Castillo-Rogez, J. C., McCord, T. B., 2010. *Icar* 205, 443

Combe, J.-P., McCord, T. B., Tosi, F., et al., 2016. *Sci* 353, aaf3010

De Sanctis, M. C., Ammannito, E., Raponi, A., et al. 2015, *Natur* 528, 241

De Sanctis, M. C., Coradini, A., Ammannito, E., et al. 2011. *SSRv* 163, 329

De Sanctis, M. C., Raponi, A., Ammannito, E., et al. 2016. *Natur* 536, 54

Green, D. W., Marsden, B. G., Morris Ch. S. 2001, *Astron. Geophys.* 42, 1.11

Hansen, J. E. 1969, *JAtS* 28, 120

Hapke, B. 2012, Theory of Reflectance and Emittance Spectroscopy (2nd ed.; Cambridge: Cambridge Univ. Press)

Holzschuch, N., Gascuel, J-D. 2013, *ICGA* 33, 66

Küppers, M., O'Rourke, L., Bocklelée-Morvan, D., et al. 2014, *Natur*, 505, 525

Li, J.-Y., Le Corre, L., Schröder, S. E., et al. 2013, Icar, 226, 1252

Li, J.-Y., McFadden, L. A., Parker, J. Wm., et al. 2006, Icar, 182, 143

Li, J.-Y., Reddy, V., Nathues, A., et al. 2016, ApJL, 817:L22

McCord, T. B., Sotin, C. 2005. *JGR* 110, E05009

Molaro, P., Lanza, A. F., Monaco L., et al. 2016. *MNRAS* 458, L54

Nathues, A., Hoffmann, M., Cloutis, E. A., et al. 2014, *Icar* 239, 222

Nathues, A., Hoffmann, M., Schaefer, M., et al. 2015, *Natur* 528, 237

Nathues, A., Platz, Th., Thangjam, G., et al. *ApJ* (in review)

Nathues, A., Hoffmann, M., Platz, T., et al. 2016, *PSS* (in press)

Park, R. S., Konopliv, A. S., Bills, B. G., et al. *Natur* 537, 515

Platz, T., Nathues, A., Schorghofer, N., et al. *Natur Astron.* (accepted)

Rivkin, A. S., campins, H., Emery J. P. 2016, Asteroids IV, 65 (University of Arizona Press)

Roth, L., Ivchenko, N., Retherford K. D., et al. 2016, *GRL* 43, doi:10.1002/2015GL067451

Russell, C. T., Raymond, C.A., Ammannito E., et al. 2016 *Sci* 353, 1008

Sierks, H., Keller, H.U., Jaumann, R., et al. *SSRv* 163, 263

Thangjam, G., Nathues, A., Mengel, K., et al., 2014. *MAPS* 49, 1831

Thomas, P. C., Parker, J. Wm., McFadden, L. A., et al. 2005, *Natur* 437, 224

Travis, B.J. 2015, Bland, P.A. 2015, Feldmann, W.C., et al. 2015, 46th LPSC, 2360

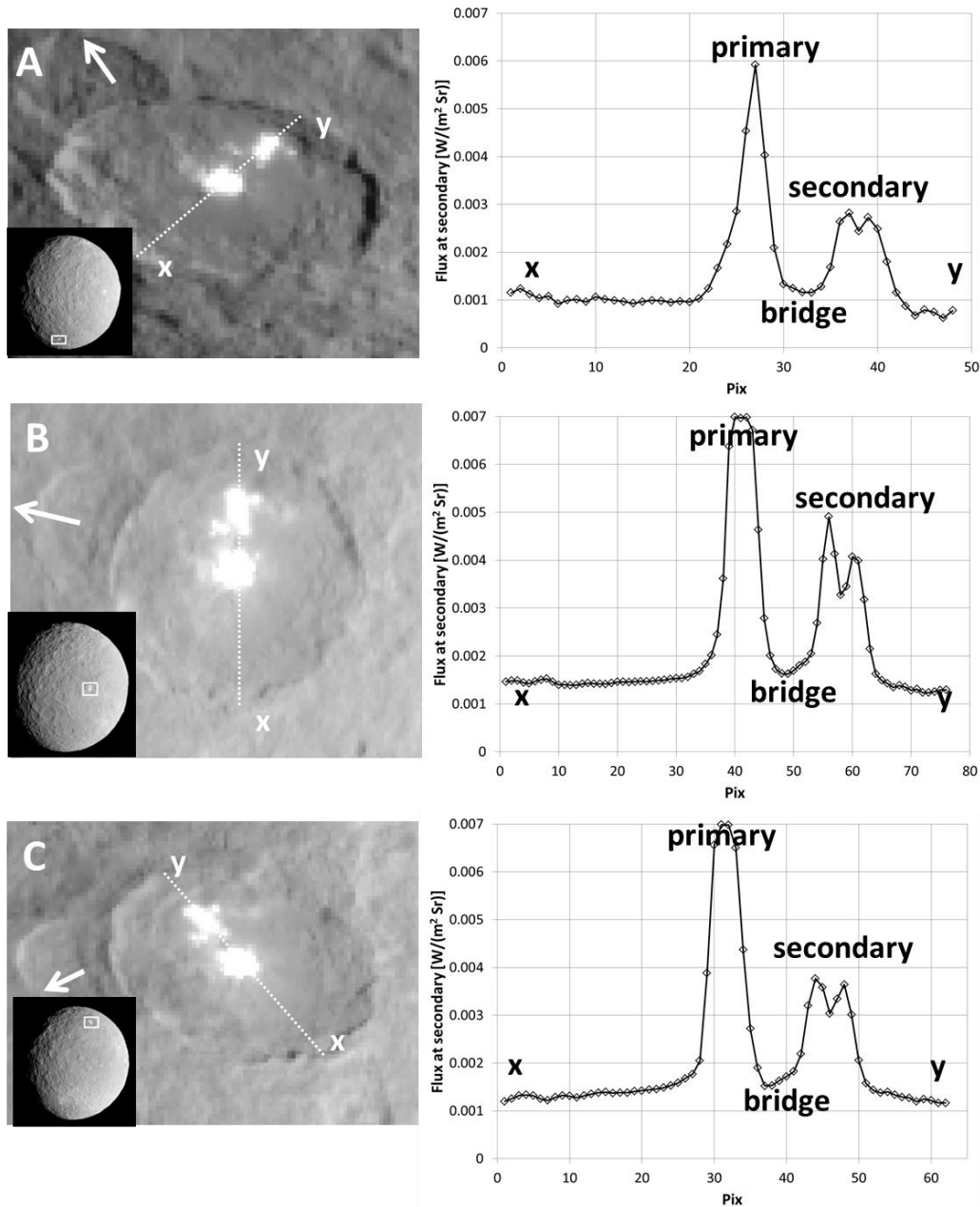

Fig. 1. Occator crater at different local solar times. Insets show the location of Occator at crater on Ceres' disk (boxes). Data have been obtained during RC3N cycle on 04-05-2015. Right subfigures display flux profiles along the displayed traces. The stretch in all images is the same. Arrows point to the north. The primary and secondary spots and the bridge are annotated in the profile. (A) FC21B0036595, local morning. (B) FC21B0036633, local noon. (C) FC21B0036662, local evening. Flux values at 'secondary' (first maxima at secondary) and 'bridge' (minima) are used for the flux ratio.

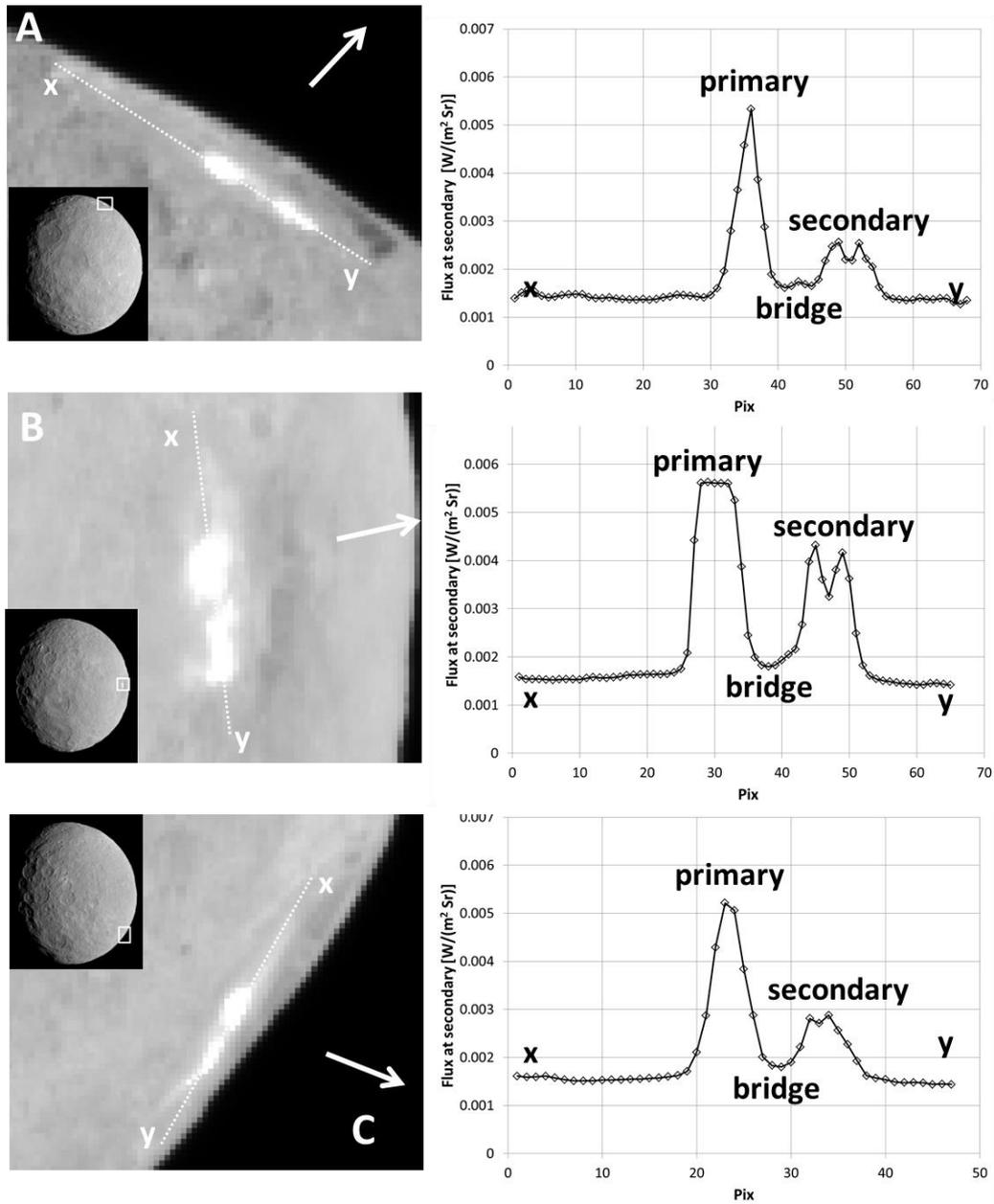

Fig. 2. Occator crater at similar local solar times as in Fig. 1, obtained during RC3S cycle on 07-05-2015. The stretch in all images is the same. Arrows point to the north. The primary and secondary spots and the bridge are annotated in the profile. (A) FC21B0037113, local morning. (B) FC21B0037142, local noon. (C) FC21B0037169, local evening. Flux values at secondary (first maxima at secondary) and bridge (minima) are used for the flux ratio.

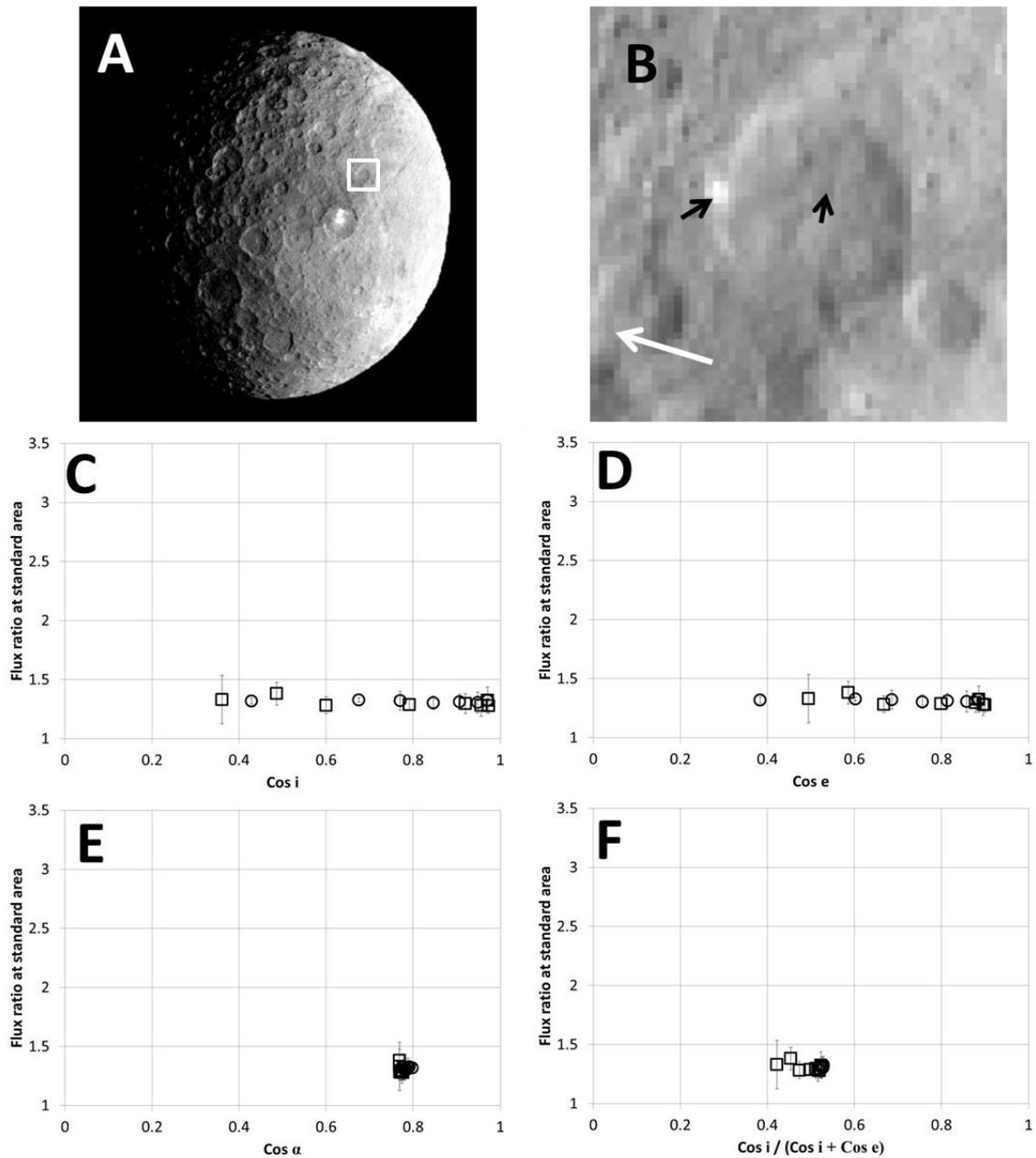

Fig. 3: (A). Image (FC21B0036634) shows the location of our reference area (box). (B) Enlarged view of the reference area. The black arrows point to the reference sites where the fluxes have been measured. The white arrow points to the north. (C-D) Flux ratios of two reference areas versus cos of incidence angle (C), cos of emission angle (D), cos of phase angle (E), and [(cos i + cos e) /cos i] where i and e denote incidence and emission angles, respectively (F). Square are for morning and circles are for afternoon.

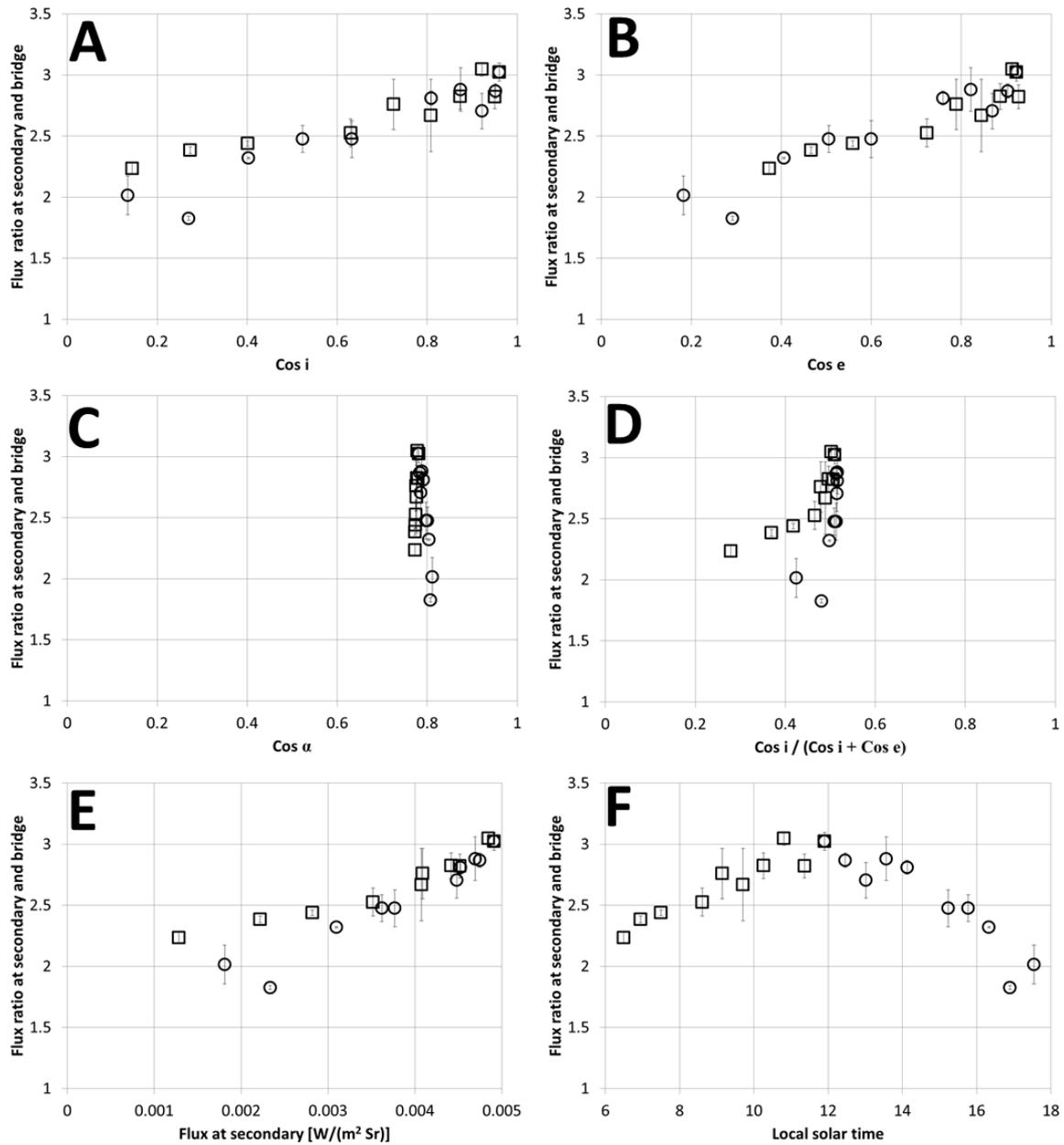

Fig. 4: (A-F) Flux ratio between 'secondary' and 'bridge' obtained from RC3N images plotted against cos of incidence angle (A), cos of emission angle (B), cos of phase angle (C), and [cos i + cos e ) /cos i] where i and e denote incidence and emission angles, respectively (D), flux at secondary (E), and local solar time (F). Square are for morning and circles are for afternoon.

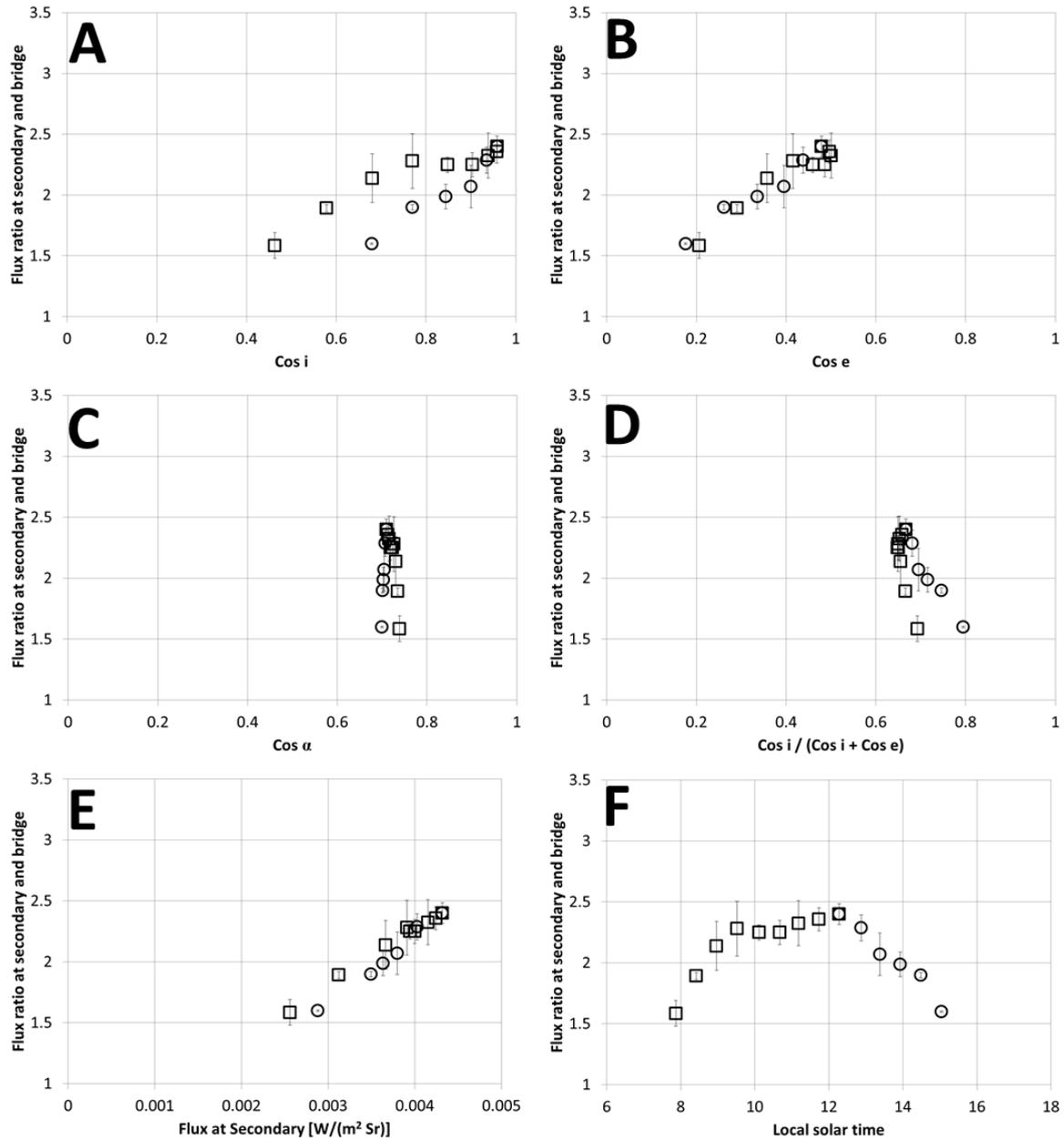

Fig. 5: (A-F) Flux ratio at 'secondary' and 'bridge' obtained from RC3S images plotted against cos of incidence angle (A), cos of emission angle (B), cos of phase angle (C), and [cos i + cos e ) /cos i] where i and e denote incidence and emission angles, respectively (D), flux at secondary (E), and local solar time (F). Square are for morning and circles are for afternoon.